\documentclass[letterpaper]{article} 
\usepackage{aaai25}  
\usepackage{times}  
\usepackage{helvet}  
\usepackage{courier}  
\usepackage[hyphens]{url}  
\usepackage{graphicx} 
\usepackage{natbib}  
\usepackage{caption} 
\usepackage{multirow}
\usepackage{booktabs}
\usepackage{dblfloatfix}

\usepackage{algorithm}
\usepackage{algorithmic}

\usepackage{newfloat}
\usepackage{listings}
\DeclareCaptionStyle{ruled}{labelfont=normalfont,labelsep=colon,strut=off} 
\floatstyle{ruled}
\newfloat{listing}{tb}{lst}{}
\floatname{listing}{Listing}
\nocopyright
\title{Towards a General Framework for HTN Modeling with LLMs}

\title{Towards a General Framework for HTN Modeling with LLMs}
\author {
    Israel Puerta-Merino\textsuperscript{\rm 1},
    Carlos Núñez-Molina\textsuperscript{\rm 2},
    Pablo Mesejo\textsuperscript{\rm 1},
        Juan Fernández-Olivares\textsuperscript{\rm 1}
}
\affiliations {
    \textsuperscript{\rm 1}University of Granada, Spain\\
    \textsuperscript{\rm 2}RWTH Aachen University, Germany\\
    israelpm01@ugr.com, carlos.nunez@ml.rwth-aachen.de, 
        pmesejo@ugr.es, faro@decsai.ugr.es
}

\begin{document}

\maketitle

\begin{abstract}

The use of Large Language Models (LLMs) for generating Automated Planning (AP) models has been widely explored; however, their application to Hierarchical Planning (HP) is still far from reaching the level of sophistication observed in non-hierarchical architectures. In this work, we try to address this gap. We present two main contributions. First, we propose L2HP, an extension of L2P (a library to LLM-driven PDDL models generation) that support HP model generation and follows a design philosophy of generality and extensibility. Second, we apply our framework to perform experiments where we compare the modeling capabilities of LLMs for AP and HP. On the PlanBench dataset, results show that parsing success is limited but comparable in both settings (around 36\%), while syntactic validity is substantially lower in the hierarchical case (1\% vs. 20\% of instances). These findings underscore the unique challenges HP presents for LLMs, highlighting the need for further research to improve the quality of generated HP models. 
\end{abstract}

%
\begin{links}
    \link{Code}{https://github.com/Corkiray/L2HP}
\end{links}

\begin{figure*}[t]
    \centering
    \includegraphics[width=\linewidth]{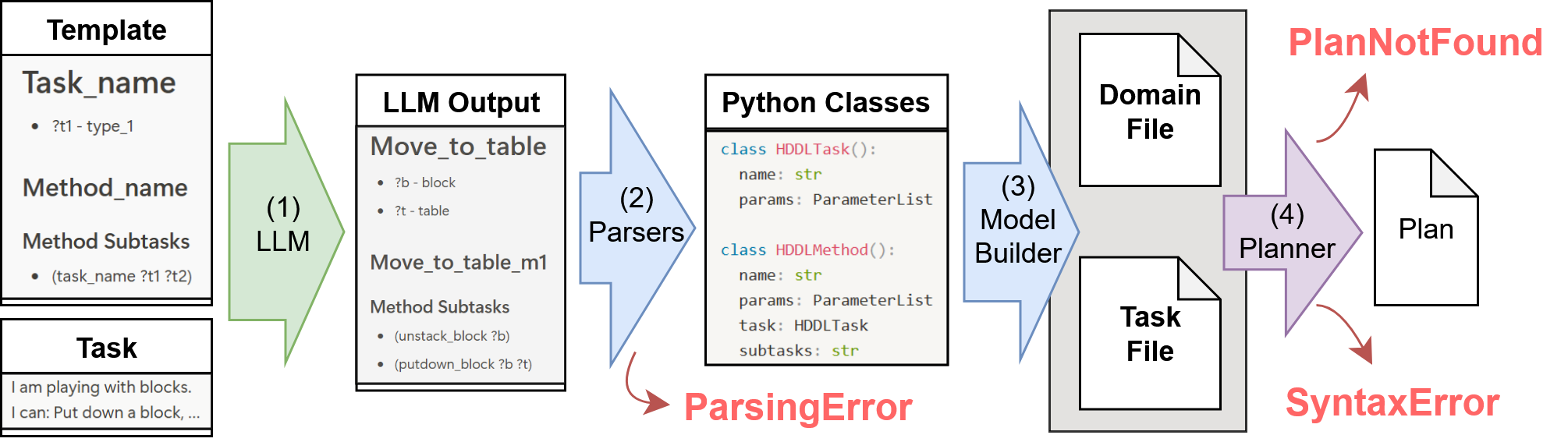}
    \caption{Overview of a typical L2HP workflow: (1) The LLM is asked to model a planning task, following a template written in Markdown. (2) The LLM output is parsed to extract the planning elements into structured Python Classes. (3) Using the structured data, the planning model is generated -- i.e., both the domain and task files are created. (4) A symbolic planner is then invoked to generate a plan that solves the given problem.}
    \label{fig:overview}
\end{figure*}

\section{Introduction}


Automated Planning (AP) is a foundational and widely applied area within AI. Despite its successes, AP continues to face two major challenges: (1) the high computational cost associated with solving large-scale problems, and (2) the difficulty of accurately formalizing complex world models. The first challenge can be mitigated through the Hierarchical Planning (HP) approach. HP leverages multiple levels of abstraction to accelerate planning and produce more interpretable, human-like plans.

Concurrently, recent advancements in AI, particularly the development of Large Language Models (LLMs), offer promising tools to address the second challenge. LLMs, trained on massive textual corpora using deep neural architectures, exhibit strong natural language (NL) understanding and generation capabilities, resulting in a great adaptability to a wide variety of contexts without requiring additional task-specific training. Their proficiency in translating information into structured outputs positions them as promising candidates for model generation (i.e. domain plus instance files generation) in AP.

Given this potential, integrating LLMs into HP systems emerges as a compelling research direction. We posit that hybrid planning frameworks, where LLMs generate hierarchical models and symbolic planners leverage these models to search for a solution plan, represent a promising avenue.

LLMs have received increasing attention for classical AP modeling \cite{tantakoun2025llmsplanningmodelerssurvey}. However, as shown by \citet{puerta2025roadmap}, there remains a substantial gap between research on the integration of in LLMs into AP and their specific integration into the HP life-cycle. To address this gap and encourage the research of LLMs in HP, we introduce two main contributions. First, we propose \textbf{L2HP}, a framework built upon the L2P library \cite{tantakoun2025llmsplanningmodelerssurvey} to support LLM-driven model generation for HP. We designed L2HP with a focus on generality and extensibility, to facilitate and encourage the implementation of future LLM-driven HP systems. Second, we apply L2HP to conduct a preliminary \textbf{empirical study} that shows the limitations of current LLM-based HP modeling systems, highlighting the need for further research into this field.


We use PlanBench \cite{valmeekam2023planbench}, a dataset designed to evaluate planning and reasoning capabilities in NL, to compare the quality of AP and HP models generated with a baseline LLM method.\footnote{\url{https://huggingface.co/datasets/tasksource/planbench}}    
Our results reveal two key findings. First, LLMs exhibit significant difficulty in adhering to the expected output structure, with over 60\% parsing failure across both AP and HP. Second, the syntactic validity is substantially lower for HP compared to AP (1\% vs. 20\% of instances, respectively). 
Our findings demonstrate the limitations of standalone LLMs for AP model generation, which accentuate in the case of HP. This highlights the need for further research into this area, for which we hope our framework will serve as a valuable tool.

\section{Background}


\subsection{Hierarchical Planning}

HP extends AP by solving problems through multiple levels of abstraction \cite{ghallab2004automated, geffner2013concise}. This approach is grounded in two foundational principles: (1) many real-world problems exhibit an inherent hierarchical structure that can be exploited to guide planning, and (2) this structure mirrors human problem-solving strategies, typically more hierarchical than sequential. Consequently, HP often results in more interpretable and computationally efficient planning than classical AP. 

The most prominent formalism of HP is the Hierarchical Task Networks (HTN) framework \cite{nau2003shop2}. In HTN planning, abstract tasks are iteratively decomposed into simpler subtasks through predefined methods, until a sequence of primitive actions is obtained. To support consistent domain modeling across various planners, several languages have been proposed, highlighting HPDL \cite{fdez2006bringingsiadex} and HDDL \cite{holler2020hddl}, both extending the classical PDDL syntax to support a smooth transition to HP structures. HDDL has become the most widely adopted standard due to its broad planner support. 

However, despite its advantages in scalability, interpretability, and knowledge reuse, HP presents its own challenges. Chief among these is the need for domain experts to define tasks, methods, and hierarchical rules, increasing the complexity of modeling, especially for new or poorly structured domains \cite{erol1994htn}. Furthermore, since it is a more expressive approach than classical AP, the modeling process, as well as the validation, debugging, and maintenance of HTN models, becomes more complex \cite{erol1994htn}. Moreover, the absence of automated tools to assist in creating or refining hierarchical models further limits adoption of HP \cite{zhuo2014learning}. Therefore, although HP aligns well with how humans decompose and reason about complex tasks, its scalability and adoption in new environments are currently limited by the difficulty of its modeling process.

\subsection{Planning with LLMs}

In last years, the use of LLMs in planning is being widely explored. This is a relatively new and rapidly evolving area of research. Early studies investigated how LLMs could assist AP by suggesting action sequences or guiding search processes. Subsequent work extended their use to translating (NL into formal models, inferring symbolic representations from observations, and even acting as approximate planners. Given the diversity and pace of this field, several surveys have emerged to consolidate and clarify the state of the art \cite{pallagani2024prospects, huang2024understanding, valmeekam2022large}. Within this context, two major lines of research can be identified: LLMs-as-Planners and LLMs-as-Modelers.

The motivation for using LLMs directly as reasoning agents or planners is reasonable. These models, trained on vast textual corpora, can perform approximate reasoning without relying on structured symbolic representations. One of the earliest works in this line is the approach of \citet{silver2022pddl}, but the field has rapidly advanced, with increasingly sophisticated architectures aimed to enhance planning capabilities. These systems incorporate techniques that has been shown to significantly improve LLMs performance, such as Chain of Thought \cite{wei2022chain}, Few-shot prompting \cite{brown2020language}, output refinement \cite{madaan2024self}, etc. One of the most works following this approach is the LLM-Modulo \cite{kambhampati2024llms}.

On the other hand, the use of LLMs as model generators is also well-motivated. Their capacity to generate structured text, encode commonsense knowledge, and operate in both natural and formal languages positions them as valuable tools for knowledge engineering in planning. LLM+P \cite{liu2023llm} is one of the first works to generate partial AP models with LLMs, by translating a task NL description into a PDDL task file. Several notable works have since extended this idea, integrating LLMs into more sophisticated architectures, as NL2Plan \cite{gestrin2024nl2plan} and \cite{guan2023leveraging}. For an in-depth overview, see the survey by \cite{tantakoun2025llmsplanningmodelerssurvey}, which focuses specifically on the LLMs-as-Modelers paradigm.

\subsection{LLMs in Hierarchical Planning}

As shown above, LLMs have received increasing attention for classical AP modeling and are being progressively integrated into the AP life cycle. However, a substantial gap remains between research on the integration of LLMs into AP in general and their specific integration into the HP life cycle, with only a limited number of studies addressing this topic \cite{puerta2025roadmap}.

Following the LLMs-as-Planners paradigm, some works have explored the use of LLMs to generate hierarchically structured plans. These plans are typically represented in NL and generated iteratively, by prompting the LLM to decompose a task into subtasks until a complete plan is formed \cite{tse2024improving, tianxing2025step, zhao2024epo, yang2024oceanplan}. As with AP, more sophisticated architectures can be developed to improve the LLM performance. For example, \cite{kienle2025hitamp} builds on the LLM-Modulo framework to generate hierarchical plans.

In the LLMs-as-Modelers line, one of the first HP approaches is \cite{luo2023obtaining}, which uses an HTN-like NL structure to generate a task representation that is then translated into a Linear Temporal Logic (LTL) model. In the specific context of HTN generation, more recent works have also emerged \cite{fine2024leveraging, sinha2024leveraging, munoz2025chathtn}.

Finally, there are also noteworthy hybrid approaches: \cite{dai2024optimal} employs an LLM as heuristic to accelerate an LTL planning process, while \cite{hsiao2025critical} combines a symbolic HTN planner for high-level reasoning and an LLM Agent as low level planner.


\subsection{L2P Library}

Although the integration of LLMs within various stages of the AP life-cycle is being widely explored, the lack of standardized tools and methodologies continues to limit the scalability and reproducibility of LLM-based planning research. At the same time significant efforts are underway to simplify LLM integration through streamlined prompt engineering, API access, and output parsing. Notable contributions include the LangGraph framework\footnote{\url{https://github.com/langchain-ai/langgraph}} and the Model Context Protocol\footnote{\url{https://github.com/modelcontextprotocol}}.

In the specific context of AP, one of the pioneering efforts is L2P  -- LLM-driven Planning Model library kit \cite{tantakoun2025llmsplanningmodelerssurvey}. To the best of our knowledge, L2P is the first library designed to facilitate the LLM-driven generation of PDDL models (i.e. domain plus instance files). Given its scope, modularity and extensibility, L2P was selected as the foundational platform for building our HP-focused extension: L2HP.

L2P offers a suite of tools for building custom architectures. Its core functionality centers on a set of \textit{extraction} methods, which allow to easily use an LLM to obtain specific structured data representations of PDDL components (e.g., actions, predicates, parameters). These methods are conformed by pipelines that apply prompt templates, invoke an LLM, and parse the generated responses into structured objects. This general workflow -- illustrated in steps (1-2) of Figure \ref{fig:overview} -- is always similar across use cases, but can be applied at varying levels of complexity, from generating individual actions to producing complete PDDL domain files.\footnote{Figure~\ref{fig:overview} represents a typical L2HP workflow, and does not exactly reflect L2P behavior (e.g., L2P does not use Markdown). However, since L2HP is inspired by  and built on top of L2P, the overall process is conceptually similar and closely reflect the core logic of L2P} This workflow proceeds as follows: 
\begin{enumerate}
    \item \textbf{Template Application}. The NL task description is embedded into a prompt template designed to elicit a specific structured response.
    \item \textbf{LLM Invocation}. The prompt is submitted to an LLM via API or local inference.
    \item \textbf{Response Parsing}. The LLM output, which is expected to match the template, is parsed into structured elements using custom parsers.
\end{enumerate}

The orange elements in Figure \ref{fig:componentsDiagram} represents the core components of the L2P library. Some of these will be discussed in more detail in the \textit{L2HP Framework} section. Briefly, L2P provides a collection of templates and parsers for various planning constructors. Parsed elements are managed using two builder classes: one for domain-level definitions and another for task-level specifications. These builders not only manage structured data but also implement the mentioned extraction methods, along with additional functionalities such as automatic PDDL generation. L2P also offers seamless integration with LLM APIs and a planner, as well as validation tools, feedback mechanisms, and a prompt construction toolkit. Its versatility is further demonstrated by its integration of several previously decentralized, state-of-the-art LLM-based planning architectures.




\section{L2HP Framework}

\begin{figure*}[ht]
    \centering
    \includegraphics[width=\linewidth]{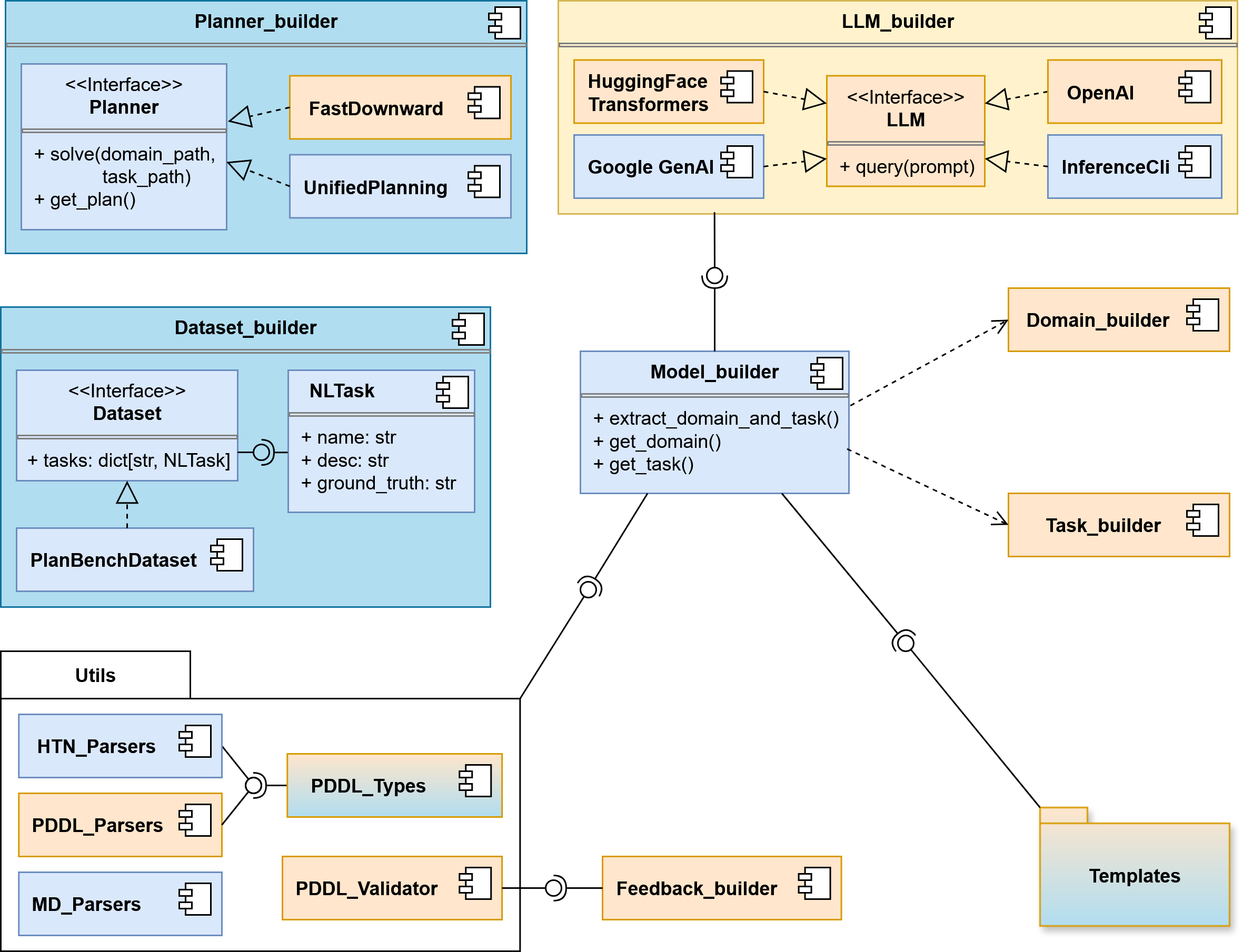}
    \caption{Component Diagram of L2HP. Orange components are native to the L2P library, while blue ones were developed within L2HP. Components with both colors indicate that they were originally part of L2P and have been extended in L2HP.}
    \label{fig:componentsDiagram}
\end{figure*}

L2HP is introduced as the first framework designed to support the use of LLMs as model generators for HP. However, its overarching goal extends beyond only reducing entry barrier and encourage the community to research in this emerging area; it also aims to simplify the development and evaluation of LLM-based planning architectures. Therefore, the design of L2HP is guided by two core principles:

\begin{enumerate}
    \item \textbf{Generalization and Extensibility}. L2HP is structured to allow seamless integration of LLMs, planners, templates and architectures, ensuring that the experiments are reproducible and independent of implementation details. 
    Note that, due to the nondeterministic nature of LLMs, the results might variate even when running the exactly same experimentation.
    \item \textbf{HP Integration}. We provide a suite of tools designed for HTN modeling and planning, including support for HPDL, HDDL, and HTN-compatible planners.
\end{enumerate}

To achieve this, L2HP builds upon L2P by extending its capabilities to support hierarchical structures. It introduces new tools and abstractions to promote experimentation, along with additional integrations specifically designed for HP. All components that conform L2HP are described in the following subsections and illustrated in Figure \ref{fig:componentsDiagram}.

\subsection{Generalization and Extensibility}

L2HP has integration with a variety of planners and LLMs with modular compatibility. To promote flexible experimentation, L2HP introduces a general \texttt{Planner} class that standardizes the execution of the different planners with a same interface. This interface currently supports the Unified Planning \cite{unified_planning_softwarex2025} library  -- highlighting the ARIES planner for HDDL -- and maintains compatibility with the Fast Downward planner interface inherited from L2P. The interface is simple, to encourage the future integration of additional planners, especially HTN-specific ones.

Following a similar principle, L2HP extends the LLM interface introduced by L2P, supporting four distinct methods for executing LLMs: the two original ones from L2P: Local inference via HuggingFace Transformers and Remote inference via OpenAI’s GPT API; and two new options: The InferenceClient API, for accessing the HuggingFace-hosted LLMs, and Google’s GenAI API for the Gemini LLMs.

Datasets integration is also streamlined through a interface conformed by a series of standardized tasks, ensuring consistency and accessibility across future experiments. Figure \ref{fig:L2HP-dataset-example} illustrates an example of an L2HP task. As an initial set of AP tasks described in NL, we have incorporated the PlanBench dataset \cite{valmeekam2023planbench}. Further details about PlanBench are provided in \textit{Experimentation} section.


L2HP encourages the use of Markdown (MD) as the default output format when building more robust and consistent templates and parsers. Therefore, L2HP provides a built-in parsing toolkit, which includes additional MD utilities to extract sections by title or convert lists into structured data. Markdown is selected for several reasons:
\begin{itemize}
    \item \textbf{Consistency.} LLMs are well-trained on MD and tend to output it more reliably than PDDL, HDDL, or another uncommon or hand-made format.
    
    \item \textbf{Performance.} MD gives the LLM space to carry out the reasoning or knowledge retrieval process that has already been seen to improve results \cite{wei2022chain}.
    
    \item \textbf{Compatibility.} MD aligns closely with the original L2P representations, facilitating reuse and extension of tools.

    \item \textbf{Extensibility.} MD offers structured formatting that is visually intuitive, making it easier to build new user-defined templates and parsers.

\end{itemize}

Finally, a new \texttt{model\_builder} class is provided. It integrates the inherit L2P domain and task builders, preserving all L2P functionalities, but they are also extended to support HTN, as will be illustrated in the next subsection. This builder also provides new functions to export the internal models into multiple formats (PDDL, HPDL, and HDDL).

\subsection{Hierarchical Planning Integration}

To extend L2P to support HTN planning, we introduce the following enhancements:
\begin{enumerate}
    \item \textbf{New Data Types}. In addition to the native L2P structures (Parameters, Predicates, Actions, etc), we provide structures for representing HTN-specific data, such as tasks and methods, both in HPDL and in HDDL syntax.
    \item \textbf{HTN Parsers}. Custom parsers that extract an store HTN elements from structured LLM output into built-in types.
    \item \textbf{Prompt Templates}. Designed to guide LLMs in producing output compatible with the built-in parsers.
\end{enumerate}

An integrated extraction method leverages these components to automatically generate complete planning models from NL tasks (steps 1, 2 and 3 of Figure \ref{fig:overview}). This method is integrated into the new model builder mentioned above. It supports both classical and hierarchical modes, employs all the tools mentioned before, and also serves as a reference for developing new pipelines.

Finally, we present NL2HTN, a simple and illustrative architecture that combine all the provided functionalities to automatically create plans from NL tasks (Figure \ref{fig:overview}). It supports both PDDL and HDDL modes, and can be instantiated with interchangeable components (e.g., LLMs, planners, templates), aligning with the framework's goal of enabling flexible experimentation.

\section{Experimentation}

To highlight the current gap between AP and HP with LLMs, we conducted preliminary experiments using the NL2HTN architecture and the PlanBench dataset. This evaluation has two additional goals: (1) to establish a baseline for LLM-based HP model generation, and (2) to demonstrate a practical use case of L2HP for empirical studies within HP.




\subsection{PlanBench Dataset}

\begin{figure}[h]
    \centering
    \includegraphics[width=0.8\linewidth]{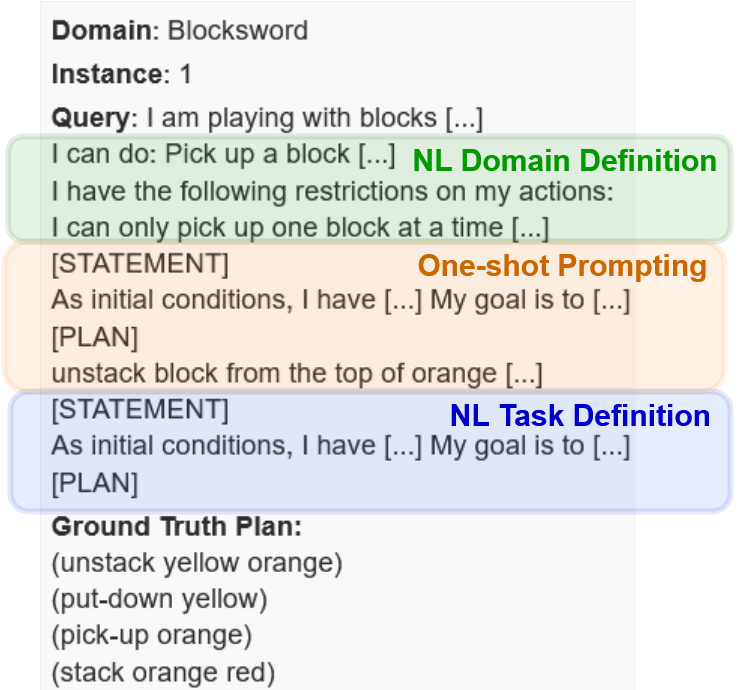}
    \caption{An instance of the PlanBench generation subset.}
    \label{fig:planbench-example}
\end{figure}

PlanBench \cite{valmeekam2023planbench} is a dataset that includes various tasks designed to evaluate planning and reasoning capabilities in NL. We focus on the plan generation subset, which includes 2270 problem instances spanning five different domains. Each instance represents an AP task, and has the following main several components: the \textbf{domain} name, a \textbf{instance} identifier, a \textbf{ground-truth plan} (i.e., a valid plan that solves the problem), and the task \textbf{query} itself. The query consists of a prompt with three distinct elements (Figure \ref{fig:planbench-example}):
\begin{enumerate}
    \item \textbf{NL Domain Definition}. Explicit NL specifications of all available actions, including their constraints. 
    \item \textbf{NL Task Definition}. A description of the initial state and the goal to archive.
    \item \textbf{One-shot Prompting}. A separate problem instance within the same domain along with a valid solution plan. This approach has been shown to significantly improve LLM performance \cite{brown2020language}.
\end{enumerate}


PlanBench is particularly well-suited for evaluating the modeling capabilities of LLMs, as it provides a diverse set of fully specified planning problems in NL. This setup primarily tests an LLM’s ability to accurately interpret detailed task descriptions and convert them into structured representations, with minimal reliance on generating additional information through knowledge retrieval. For this reason, we consider PlanBench the most appropriate benchmark to establish a starting point for research within LLM modeling capabilities. Consequently, it has been integrated into L2HP.

\begin{figure}[b]
    \centering
    \includegraphics[width=0.8\linewidth]{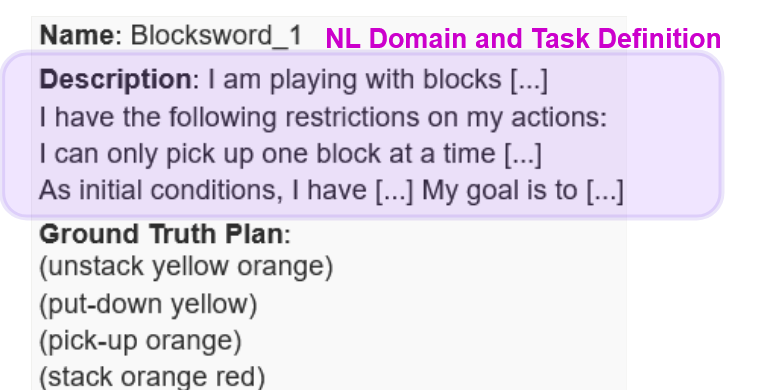}
    \caption{An instance of a standardized task. The task shown is the same as in Figure \ref{fig:planbench-example}, after being preprocessed to conform to the L2HP standardized dataset structure.}
    \label{fig:L2HP-dataset-example}
\end{figure}

We integrate PlanBench with a focus on the framework’s principles of modularity and consistency, aiming to facilitate future extension and integration with other datasets. To this end, we define a standardized dataset structure, into which each PlanBench instance is preprocessed. This structure (Figure \ref{fig:L2HP-dataset-example}) is conformed by three key elements: (1) the \textbf{task name}, (2) the \textbf{ground-truth plan} for evaluation, and (3) the \textbf{NL description} of the planning task. Therefore, the domain and problem descriptions are merged into a single task description, while the resolution example is omitted. 

\begin{figure*}[t]
    \centering
    \includegraphics[width=\linewidth]{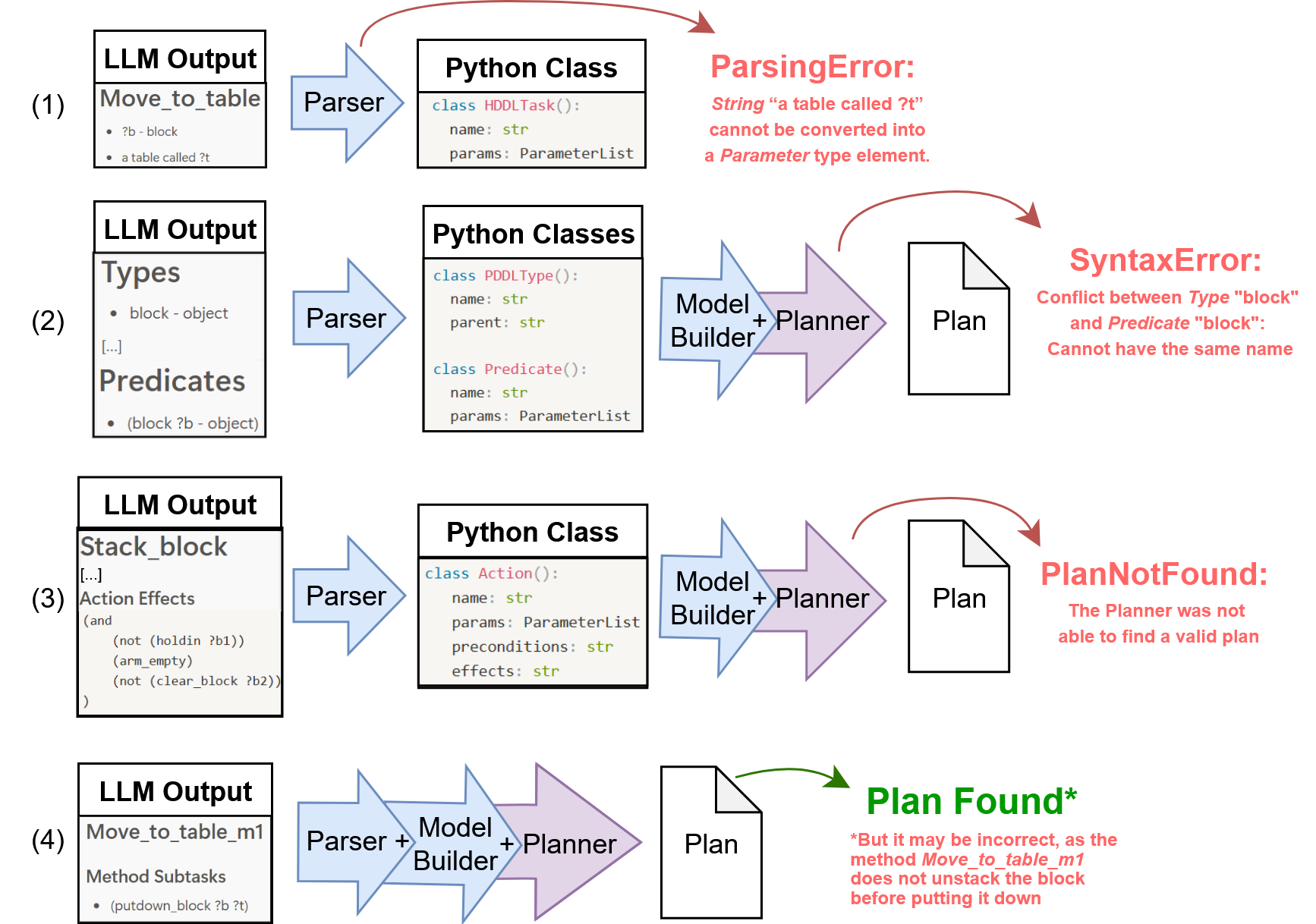}
    \caption{Illustrative examples of different types of invalid LLM outputs. Examples (1) to (3) represents the three core categories of identifiable errors, while (4) shows an incorrect output that does not trigger an error assertion. In (1), the phrase \texttt{a table called ?t} lacks a proper structure, so the parser is unable to translate it into a valid parameter. In (2), both the object \texttt{block} and the type \texttt{block} are parsable, but a conflict arises when invoking the planner. In (3), no plan can be found because the Stack\_block action lacks the \texttt{(on ?b2 ?b1)} effect. In (4), the planner finds a plan; however, it is likely incomplete, as the method Move\_to\_table\_m1 lacks the \texttt{(unstack\_block ?b)} subtask. A correct definition is shown in Figure \ref{fig:overview}.}
    \label{fig:error-examples}
\end{figure*}



\begin{table*}[h]
\centering
\begin{tabular}{llclrrrlrrr}
                                    &  & \multicolumn{1}{l}{} &  & \multicolumn{3}{c}{\textbf{PDDL}}                                        &  & \multicolumn{3}{c}{\textbf{HDDL}}                                        \\
\multicolumn{1}{c}{\textbf{Domain}} &  & N                    &  & \multicolumn{1}{c}{SP} & \multicolumn{1}{c}{VS} & \multicolumn{1}{c}{S} &  & \multicolumn{1}{c}{SP} & \multicolumn{1}{c}{VS} & \multicolumn{1}{c}{S} \\ \hline \noalign{\smallskip}
\textit{Blocksworld}                &  & 600                  &  & 99                     & 92                     & 88                     &  & 0                      & 0                      & 0                      \\
\textit{Mystery}                    &  & 600                  &  & 169                    & 121                    & 98                     &  & 423                    & 16                     & 10                     \\
\textit{Depots}                     &  & 500                  &  & 237                    & 78                     & 64                     &  & 343                    & 6                      & 4                      \\
\textit{Logistics}                  &  & 285                  &  & 173                    & 102                    & 85                     &  & 29                     & 1                      & 0                      \\
\textit{OD\_ Logistics}             &  & 285                  &  & 79                     & 68                     & 66                     &  & 108                    & 7                      & 3                      \\ \hline \noalign{\smallskip}
\textbf{Total}                      &  & 2270                 &  & 757                    & 461                    & 401                    &  & 903                    & 30                     & 17                     \\ \hline \noalign{\smallskip}
\multicolumn{3}{c}{\textbf{Average:}}                         &  & \textbf{33.35\%}       & \textbf{20.31\%}       & \textbf{17.67\%}       &  & \textbf{39.78\%}       & \textbf{1.32\%}        & \textbf{0.75\%} 
          \\ \hline \noalign{\smallskip}
\end{tabular}
\caption{Summary of the execution of NL2HTN through the PlanBench plan generation subset. Showing the number of problems (N) per domain and the results of NL2HTN in each mode (PDDL and HDDL). The scores shown are explained in the \textit{Experimentation} section, representing the number of LLM outputs successfully parsed (SP), with a valid syntax (VS), and solvable (S). The final row reports average scores across all instances. Note that each metric is a prerequisite for the next, leading to progressively lower passing rates.}
\label{tab:my-table}
\end{table*}

\subsection{Experimental Considerations}

When executing the NL2HTN pipeline (Figure \ref{fig:overview}), various types of invalid LLM outputs can arise -- each of which should be identify and, ideally, corrected or prevented. Based on the type of error that they generate, we can categorize these invalid outputs into four groups (Figure \ref{fig:error-examples}):
\begin{enumerate}
    \item \textbf{Parsing Error.} A parser is unable to build the internal representation of some element. This occurs when output includes noise, incoherent fragments or even valid content but presented in an incorrect structure (Figure \ref{fig:error-examples}(1)).
    
    \item \textbf{Syntax Error.} The model is parsable, but the planner is unable to load it and initiate the search process. This may result from structural inconsistencies that were not detected by the parsers, or from conflicts between different element definitions (Figure \ref{fig:error-examples}(2)).
    
    \item \textbf{Plan Not Found.} The planner is able to initiate a search but fails to find a plan. This indicates that the model is syntactically valid but logically incorrect -- it does not correspond to the intended task, nor to any solvable task (Figure \ref{fig:error-examples}(3)). While this could theoretically occur due to computational limitations, our experiments use feasible, state-of-the-art planning domains, making such failures unlikely in practice.

    \item \textbf{Incorrect Plan Found.} The planner finds a plan, but it is incorrect. This occurs when the generated model is valid and coherent but does not corresponds to the intended task, or when it contains semantic errors undetectable by either the parser or planner (Figure \ref{fig:error-examples}(4)).
\end{enumerate}

While the first three error types are relatively straightforward to detect, the last presents a non-trivial challenge. Generating a plan that differs from the ground-truth does not imply that the model is incorrect, as it may describe an equivalent representation of the same task, with a different action structure. Conversely, reproducing the ground-truth does not guarantee correctness either, as the model might describe a different task that coincidentally yields the same solution. One might argue that the core problem is the lack of ground-truth models for direct comparison. However, even having such references, it would remain unresolved. A generated model could still represent a valid alternative formulation of the same problem. This highlights a deeper challenge that deserves further investigation: determining whether a generated model faithfully captures the intended problem.

With this limitation in mind, in this preliminary work we focus on the quantifiable aspects of our evaluation. The objective of this experimentation is to establish a baseline, trace the NL2HTN pipeline (Figure \ref{fig:overview}) and identify its weakness, focusing on the three core errors types that we can find within the execution. Rather than scoring outputs based directly on the error types, we evaluate each execution using three critical checkpoints. While similar, these metrics provide a more informative and interpretable perspective on the execution results:
\begin{enumerate}
    \item \textbf{Successfully Parsed (SP).} The parsers extracted all required elements from the LLM output. This indicates that the LLM output adhered to the expected structure, but it may not represent a valid task.
    \item \textbf{Valid Syntax (VS).} The LLM generated a valid planning model. However, this model may not correspond to the input task or even be a solvable one.
    \item \textbf{Solvable (S).} The planner successfully found a plan, indicating that the generated task is logically coherent. However, it may still be semantically incorrect or it may encode a different task than the one provided as input.
\end{enumerate}

We executed NL2HTN across the entire PlanBench subset in both classical (PDDL) and hierarchical (HDDL) planning modes. The specific configuration used was as follows: (LLM) \textit{Gemini 2.0 flash}, via Google GenAI; (planner) \textit{Aries}, through Unified Planning; and (templates) the two hand-made templates available in the \texttt{templates/model\_templates} directory within the L2HP repository. 


\subsection{Results}

Table~\ref{tab:my-table} summarizes the experimental results. Key findings for each of the three checkpoints are highlighted below:

\paragraph{Successfully Parsed (SP).} The percentage of SP tasks was similar in both modes -- 33\% for PDDL and 40\% for HDDL. This suggests that the ability of the LLM to follow structural prompts is relatively independent of the underlying planning paradigm. However, parsing errors were the most common failure across both planning modes, affecting more than 60\% of all tasks. This highlights the challenges that LLMs face in adhering to strict structure requirements, even when guided by well-constructed templates. 

\paragraph{Valid Syntax (VS).}  A notable disparity exists in the proportion of VS models between classical and hierarchical planning. In PDDL, most SP models were also VS -- 20\% of all tasks (60\% of the 33\% SP). By contrast, in HDDL only 1.3\% of all tasks (around a 3\% of the 40\% SP) were syntactically valid. This discrepancy is likely due to the nature of the PlanBench dataset, which provides explicit definitions for classical planning elements but lacks hierarchical annotations. Consequently, the LLM must infer the entire HTN structure from scratch, a task that requires not only syntactic precision but also domain-level reasoning. Although this requirement does not significantly affect structural consistency, as seen in the last paragraph, it has a substantial impact on the capacities of the LLM to generate valid models. Nevertheless, these findings highlight a significant limitation in the current capabilities of LLMs to infer and construct well-formed hierarchical decompositions.

\paragraph{Solvable (S).} Most VS models also generate solvable plans. In PDDL, 17\% of all tasks were S (85\% of the 20\% SV). In HDDL, 0.75\% of all tasks were S (around 60\% of the 1.3\% VS). These results highlight the potential of LLMs to generate effective planning models, as long as consistency issues are further addressed. Nevertheless, it is important to remember that finding a valid plan does not necessarily imply that the generated model corresponds to the intended task, as was exposed in the \textit{Experimentation} section.

\section{Discussions and Future Works}

The results reveal several challenges in using LLMs to model AP problems -- particularly in the context of HP. LLMs struggle with adhering to strict structures and inferring well-formed hierarchical decompositions. Additionally, a core issue in this area remains: the difficulty of automatically verify whether a generated model faithfully represents the intended task. Below, we outline promising directions and strategies to address these limitations.

\subsection{Inference Capability of LLMs} 

This is a general limitation that affects many LLM integration areas, not just AP or HP. Therefore, multiple techniques has been developed to enhance LLM reasoning. These includes the already mentioned Chain of Thoughts, few-shot prompting and output refinement, as well as additional strategies such as Tree of Thoughts \cite{yao2023tree}, best-of-k sampling, iterative generation, etc. As discussed in the \textit{Background} section, many of these techniques have already been explored in AP. However, there exists still a gap in their application into HP. Therefore, we argue that evaluating and adapting these techniques to HP -- alongside their integration into the L2HP framework -- is a valuable and necessary direction for future research.

\subsubsection{Is Fine-Tuning LLMs Worth It?}

While fine-tune can be a powerful strategy, we argue it is not the most effective approach for HTN planning. Fine-tuning a LLM requires large task-specific datasets and is both computationally and temporally expensive. Moreover, given the rapid advancement of LLMs development, fine-tuned models risk becoming obsolete quickly. Therefore, leveraging advanced prompting techniques with state-of-the-art models offers a more sustainable and cost-effective direction for this domain.

\subsection{Structural Consistency of LLMs Output}

As with inference, structural consistency is a widespread challenge. It is fundamental to the agentic LLMs area -- i.e. LLMs capable of interacting with external tools such as web APIs, databases, or code execution. In this context, some tools such as LangGraph and the Model Context Protocol (mentioned in the \textit{Background} section) have been developed to guide LLMs toward producing consistent executable outputs. In summary, these tools typically combine structured prompting with post-processing pipelines to ensure that the generated outputs are both valid and actionable. We believe that integrating such tools into L2HP, alongside the development of more robust HTN-specific parsing mechanisms, represents a high-valuable direction for future work.

Another promising approach is model repair, where flawed outputs are automatically refined based on feedback. This can be reached by incorporating into the prompt the generated error messages, feedback from some validation tool, or even responses from a secondary LLM. This mechanism  is already implemented in the original L2P framework, so extending it to support HTN models would be a natural and impactful improvement.


\subsection{Correctness of Generated Models}

As discussed in the \textit{Experimentation} section, assessing the semantic correctness of generated models remains an open challenge. While a definitive solution is still elusive, two promising directions include: (1) developing HTN-specific benchmarks that pair NL descriptions with ground-truth HTN models, and (2) embedding the architectures within interactive agents that operate in simulated or game environments, enabling a more qualitative evaluation in which correctness is inferred from the agent's behavior.

\subsubsection{Is PlanBench a Suitable Choice?}

PlanBench is one of the few datasets specifically designed for LLM-based planning, and provides a valuable set of AP tasks described in NL. While other datasets exist, most focus on another specific planning-related tasks (e.g. repairing or predicting missing actions) or are derived from PlanBench. Therefore, we consider PlanBench the most suitable choice at the present. 

However, we found notable limitations when applying it for HP. The tasks do not provide decompositions information, forcing the LLMs to infer hierarchical structures from flat AP tasks. In addition, the LLM must transform an AP domain into an HTN one, which often results in shallow or flat hierarchies, instead of fully exploiting HP’s strengths. Moreover, goals are expressed as sets of conditions, requiring the LLM to reinterpret them into equivalent HTN task formulations rather than directly translating them from NL. These issues highlight the need for new datasets or extensions of PlanBench specifically tailored for HP.

\section{Conclusion}

This work explored the applicability gap in the intersection of LLMs and HP. First, we introduced \textbf{L2HP}, an extension of the L2P library designed to encourage the LLM integration across the HP life-cycle. Second, we presented a preliminary study of LLM capabilities for model generation in both AP and HP. Beyond benchmarking, our experiments illustrates how L2HP simplifies the development and evaluation of LLM-based planning architectures. We hope that, by offering this modular and extensible framework, L2HP will serve as a valuable tool for the community and encourage further integration between LLMs and HP.


Our results show that LLMs struggle in adhering to strict structural requirements, even when guided by well-designed templates, as parsing errors were the most common failure in both planning modes. Moreover, HP presents further difficulties, being syntactic validity remarkably lower than in AP. Nevertheless, in both settings, most valid models were also solvable, suggesting that -- while further research is required -- this remains a promising direction.

From these findings, we identify three key research directions: \textbf{(1) developing HTN-specific evaluation methods}, such as dedicated benchmarks or embedded evaluation frameworks; \textbf{(2) improving output reliability} through advanced prompting techniques, and integrating them into the L2HP framework; \textbf{(3) enhance HP modeling capabilities} through more robust, specific parsers and novel techniques that boost LLM performance in constructing HP problems.

\section{Acknowledgment}

This work has been partially funded by the Grant PID2022-142976OB-I00, funded by MICIU/AEI/ 10.13039/501100011033 and by “ERDF/EU”.

\bibliography{aaai25}

\end{document}